\documentstyle[11pt,emulateapj,psfig]{article}

\def\ic{IC~1613}
\def\hii{H {\small II}}
\def\hi{H {\small I}}

\begin{document}

\title{Stellar Populations at the Center of IC 1613$^1$}

\author{Andrew A. Cole,\altaffilmark{2}
Eline Tolstoy,\altaffilmark{3}
John S. Gallagher, III,\altaffilmark{2}
John G. Hoessel,\altaffilmark{2}
Jeremy R. Mould,\altaffilmark{4}
Jon A. Holtzman, \altaffilmark{5}
Abhijit Saha, \altaffilmark{6}
Gilda E. Ballester,\altaffilmark{7} 
Christopher J. Burrows,\altaffilmark{8,9}
John T. Clarke,\altaffilmark{7}
David Crisp,\altaffilmark{10}
Richard E. Griffiths,\altaffilmark{11}
Carl J. Grillmair,\altaffilmark{12}
Jeff J. Hester,\altaffilmark{13}
John E. Krist, \altaffilmark{8}
Vikki Meadows, \altaffilmark{10}
Paul A. Scowen,\altaffilmark{13}
Karl R. Stapelfeldt,\altaffilmark{10}
John T. Trauger,\altaffilmark{10}
Alan M. Watson,\altaffilmark{14}
and James R. Westphal\altaffilmark{15}}

\altaffiltext{1}{Based on observations made with the NASA/ESA
	{\it Hubble Space Telescope}, obtained at the Space Telescope
	Science Institute, which is operated by the Association of
	Universities for Research in Astronomy, Inc., under NASA
	contract NAS 5-26555, and with the WIYN telescope at the Kitt
	Peak National Observatory.}
\altaffiltext{2}{Department of Astronomy, University of Wisconsin-Madison,
	475 N. Charter Street, Madison, WI 53706; {\it cole@astro.wisc.edu, 
	hoessel@astro.wisc.edu, jsg@tiger.astro.wisc.edu}.}
\altaffiltext{3}{European Southern Observatory, Karl-Schwarzschild-Stra\ss e,
	Garching bei M\"{u}nchen, Germany; {\it etolstoy@eso.org}.}
\altaffiltext{4}{Mt. Stromlo and Siding Spring Observatories,
	Australian National University, 
	Private Bag, Weston Creek Post Office, ACT 2611, Australia.}
\altaffiltext{5}{Department of Astronomy, New Mexico State University,
	Box 30001 Department 4500, Las Cruces, NM 88003-8001.}
\altaffiltext{6}{National Optical Astronomy Observatories, 950 
	North Cherry Avenue, Tucson, AZ 85726.}
\altaffiltext{7}{Department of Atmospheric and Oceanic Sciences,
	University of Michigan, 2455 Hayward, Ann Arbor, MI 48109.}
\altaffiltext{8}{Space Telescope Science Institute, 3700 San Martin 
	Drive, Baltimore, MD 21218.}
\altaffiltext{9}{Astrophysics Division, Space Science Department,
	European Space Agency.}
\altaffiltext{10}{Jet Propulsion Laboratory, 4800 Oak Grove Drive, 
	Mail Stop 183-900, Pasadena, CA 91109.}
\altaffiltext{11}{Carnegie-Mellon University, Department of Physics,
	5000 Forbes Avenue, Pittsburgh, PA 15213.}
\altaffiltext{12}{SIRTF Science Center, California Institute of 
	Technology, 770 South Wilson, Pasadena, CA 91125.}
\altaffiltext{13}{Department of Physics and Astronomy, Arizona
	State University, Tyler Mall, Tempe, AZ 85287.}
\altaffiltext{14}{Instituto de Astronom\'{\i}a UNAM, 58090 Morelia,
	Michoacan, Mexico.}
\altaffiltext{15}{Division of Geological and Planetary Sciences, 
	MS 170-25, California Institute of Technology, Pasadena, CA 91125.}

\setcounter{footnote}{0}

\begin{abstract}
We have observed the center of the Local Group dwarf irregular
galaxy \ic\ with WFPC2 aboard the {\sl Hubble Space Telescope}
in the F439W, F555W, and F814W filters.  We analyze the resulting
color-magnitude diagrams (CMDs) using the main-sequence and giant branch 
luminosity functions and comparisons to theoretical stellar models
to derive a preliminary star-formation history for this galaxy. 
We find a dominant old stellar population (aged $\approx$ 7 Gyr),
identifiable by the strong red giant branch (RGB) and red
clump populations. 
From the (V$-$I) color of the RGB, we estimate a mean metallicity of
the intermediate-age stellar population [Fe/H] = $-$1.38 $\pm$ 0.31.
We confirm a distance of 715 $\pm$40 kpc using the I-magnitude
of the RGB tip.  
The main-sequence luminosity function down to I $\approx$ 25 
provides evidence for a roughly constant SFR of approximately
3.5 $\times$ 10$^{-4}$ M$_{\sun}$ yr$^{-1}$ across the WFPC2 field
of view (0.22 kpc$^2$) during the past 250--350 Myr.
Structure in the blue loop luminosity function implies that
the SFR was $\approx$50\% higher 400--900 Myr ago than today.
The mean heavy element abundance of these young stars is $\approx$ 1/10 solar.
The best explanation for a red spur on the main-sequence 
at I $\approx$ 24.7 is the blue horizontal branch component
of a very old stellar population at the center of \ic.
We have also imaged a broader area of \ic\ using the 3.5-meter
WIYN telescope under excellent seeing conditions.  
The WIYN color-magnitude diagram reveals a prominent
sequence of asymptotic giant branch stars and red supergiants
that is less prominent in the WFPC2 CMD due to the smaller field
of view.  The AGB-star luminosity function is consistent with
a period of continuous star formation over 
at least the age range 2--10 Gyr.
We present an approximate age-metallicity relation for \ic, which appears
similar to that of the Small Magellanic Cloud.
We compare the Hess diagram of \ic\ to similar data for three 
other Local Group dwarf galaxies, and find that \ic\ most closely resembles
the nearby, transition-type dwarf galaxy Pegasus (DDO 216).
\end{abstract}


\section{Introduction}

The study of stellar populations is a way for astronomers to
explore galaxy evolution, providing direct information on the
variations of star-formation rate and mean abundance with time,
from the present-day back to the initial formation of a galaxy.
The ensemble of Local Group dwarf galaxies, because of their diversity of
sizes, environments, and chemical abundances
makes a good laboratory in which to investigate the processes which
control galaxy evolution (e.g., Hodge 1989; Da Costa 1998; 
Grebel 1998; Tolstoy 1999).  

Dwarf galaxies are the most common class of galaxy in the Universe,
but there are a number of unresolved questions of dwarf galaxy
evolution; the relation between the early- (elliptical/spheroidal)
and late- (irregular) type galaxies, the effects of environment
on galaxy evolution, and the patterns of chemical enrichment are
unanswered questions (Mateo 1998).  Additionally, it may be that
the ubiquitous faint, blue galaxies seen in redshift surveys 
(e.g. Ellis 1997) are the precursors of today's dwarf galaxies.
The most straightforward way to trace this possible cosmological
connection is by a determination of the star-formation histories
of the nearby dwarf galaxies (Tolstoy 1998).

Color-magnitude diagrams which include one or more main-sequence
turnoffs are a powerful tool for the determination of detailed
star-formation histories for galaxies.  However, stellar crowding
at the level of the horizontal branch and below has made this
task all but impossible for ground-based telescopes.  Consequently,
most studies to date have concentrated on the sparse, late-type
dwarf galaxies, or the relatively uncrowded outer fields around
dwarf irregulars.
WFPC2 has made it possible to push stellar photometry down below the
magnitude of the horizontal branch throughout the Local Group, 
opening new vistas for the study of stellar populations in dwarf
irregular galaxies.

In a Cycle 6 Guaranteed Time Observer program, we obtained
deep BVI images of \ic\ (DDO 8), one of the charter members of 
the Local Group (Hubble 1936), in order to constrain the ages and
metallicities of the stellar populations in this object.  In this
paper we present the WFPC2 VI color-magnitude diagram of \ic, 
complemented by ground-based images taken at the WIYN
telescope.\footnote{The
WIYN Observatory is a joint facility of the University of
Wisconsin-Madison, Indiana University, Yale University, and the
National Optical Astronomy Observatories.}
Section \ref{back} gives a summary of \ic's properties, and in 
section \ref{dat} we describe our observations and measurement
procedures.  Section \ref{cmd} presents our color-magnitude 
diagrams of \ic, and the stellar populations analysis is given
in section \ref{pop}.  We discuss \ic's range of metal abundance
and relation to a sample of Local Group dwarfs for which WFPC2
stellar population studies have been made in section \ref{disc}, and
summarize our results in section \ref{sum}.

\section{\ic: Background}
\label{back}

\ic\ was discovered by Wolf in 1906, who noted an extended grouping
of faint nebulosities near the star 26 Ceti (Wolf 1907).  \ic\ was
first resolved into stars, at magnitude 17--18, by Baade (1928),
who identified it as a Magellanic Cloud-type galaxy and compared it
to NGC 6822.  \ic\ was Baade's (1963) prime example of the eponymous
sheet of Population II red giant stars that seem to exist in all
Local Group dwarfs (Baade 1963; Sandage 1971b).

Due to low values of both foreground and
internal reddening along \ic\ sightlines, it has played an important
role in the calibration of the Cepheid variable Period-Luminosity
relation for the determination of extragalactic distances (Baade
1935, Sandage 1971a).  
\ic\ is the {\it only} Local Group dwarf irregular galaxy, apart 
from the Magellanic Clouds, in which RR Lyrae type variables have
been detected (Saha {\it et al.} 1992; Mateo 1998).

\ic\ occupies a relatively isolated position in the Local 
Group, $\approx$400 kpc from M33, $\approx$500 kpc from M31, and 
$\approx$700 kpc from the Milky Way.   Its known neighbors within
400 kpc are the comparably luminous WLM irregular
(DDO 221), at $\approx$390 kpc,
the smaller Pegasus dwarf (DDO 216), some 360 kpc distant, and the
tiny (M$_{\mathrm V}$ = $-$9.7) Pisces dwarf (LGS 3), 
at $\approx$280 kpc distance.  This minor quartet of galaxies share
similar ranges of chemical abundance, and a roughly constant long-term
star-formation rate.  They show varying amounts of recent star formation,
but each have stellar populations that in the mean are older than 
$\gtrsim$ 3 Gyr
(Pegasus--- Gallagher {\it et al.} 1998; Pisces---
Aparicio, Gallart \& Bertelli 1997; WLM--- Minniti \& Zijlstra 1996;
\ic--- Freedman 1988a; this paper).  Pegasus and Pisces have been
classified as ``transition'' galaxies, intermediate between the
irregular and spheroidal classes (Mateo 1998), although the significance
of this distinction has been called into question (Aparicio {\it et al.}
1997).  The newly discovered Andromeda VI (Pegasus dwarf spheroidal)
also lies $\approx$ 400 $\pm$40 kpc from \ic\ (Grebel \& Guhathakurta 1999).

\begin{deluxetable}{lcl}
\tablewidth{0pt}
\tablenum{1}
\tablecaption{Basic Properties of IC 1613}
\tablehead{
\colhead{Property} &
\colhead{Value} &
\colhead{Source}}
\startdata
$\ell$, b (2000.0)& 129$\fdg$8, $-$60$\fdg$6 & Mateo 1998 \nl
(m-M)$_0$         & 24.27 $\pm$0.10 & Lee {\it et al.} 1993 \nl
E$_{\mathrm B-V}$ & 0.03 $\pm$0.02  & Mateo 1998 \nl
M$_{\mathrm V}$   & $-$15.2 $\pm$0.2 & Hodge 1978 \nl
B$-$V$_0$         & 0.71 $\pm$0.1    & Hodge 1978 \nl 
M(H{\small I}) (10$^6$ M$_{\sun}$) & 54 $\pm$ 11 & Lake \& Skillman 1989 \nl
M(H{\small I})/L$_{\mathrm B}$ (M$_{\sun}$/L$_{\sun}$) & 0.81 & Mateo 1998 \nl
R$_{core}$ (arcsec)\tablenotemark{a} & 200: & Hodge {\it et al.} 1991 \nl
L (H$\alpha$) (erg s$^{-1}$) & 3.18 $\times$ 10$^{38}$ & Hunter {\it et al.}
 1993 \nl
SFR$_0$ (10$^{-3}$ M$_{\sun}$/yr)\tablenotemark{b} & 2.9 $\pm$ 0.7 & this
 paper \nl
\enddata
\tablenotetext{a}{core radius of best fit King model to unresolved light
 profile.}
\tablenotetext{b}{derived from extinction-corrected H$\alpha$ flux, see text.}
\end{deluxetable}

The basic parameters of \ic\ are summarized in Table 1.   The 
distance is unusually secure due to the large number
of Cepheid variables and low reddening 
towards \ic\ (Sandage 1971a, Freedman 1988b); the RR Lyrae distance
(Saha {\it et al.} 1992) and RGB tip distance (Freedman 1988a;
Lee {\it et al.} 1993) are in agreement to within the errors.  The
reddening maps of Burstein \& Heiles (1982) show E$_{\mathrm B-V}$
$\leq$ 0.03 mag towards \ic; Freedman (1988a) derived
E$_{\mathrm B-V}$ = 0.04.  Recent work by Schlegel, Finkbeiner
\& Davis (1998) finds a foreground component E$_{\mathrm B-V}$ = 0.01.

We include the scale length of \ic's surface brightness 
profile in Table 1, to emphasize the fact that the WFPC2
field, $\approx$150$\arcsec$ across, images only a small
portion of the central galaxy.  As Hodge {\it et al.} (1991)
emphasize, the tabulated scale length is highly uncertain
due to the low light levels, small number statistics, and 
the hard-to-define tradeoffs between model parameters.

The H$\alpha$ luminosity in Table 1 is taken from Hunter, Hawley,
\& Gallagher (1993), adjusted downwards to compensate for their 
assumed distance of 900 kpc to \ic.  The H$\alpha$ luminosity of
a galaxy provides a method to estimate the recent (0--10 Myr) star-formation
rate.  Hunter \& Gallagher (1986) derive

\begin{equation}
\mathrm \dot{M} = 7.07 \times 10^{-42}\; \mathrm L(H\alpha)\;\:
 \mathrm M_{\sun}\; \mathrm yr^{-1},
\end{equation}

\noindent with a possible additional factor of $\sim$1.6 to be added 
to account for internal extinction within their sample of calibrating 
galaxies.  We find a total star-formation rate over the past 10 Myr
of 2.9 $\pm$0.7 $\times$ 10$^{-3}$ M$_{\sun}$ yr$^{-1}$ for \ic.

\section{The Data}
\label{dat}

\subsection{Observations \& Reductions}

\ic\ was observed with WFPC2 over 9 orbits on 26--27 August, 1997.
The PC chip was centered 97$\arcsec$\ southwest of the center of 
the galaxy, at right ascension 1$\mathrm ^h$
04$\mathrm ^m$ 48\fs7, declination $+$02$\arcdeg$
07$\arcmin$ 06$\farcs$2 (J2000.0).  Exposures totalling 10,700 sec each
were taken through the F555W (WFPC2 V) and F814W (WFPC2 I)
filters.  F439W (WFPC2 B) images totalling 2,600 sec were also
obtained.  The short, blue exposures were intended to provide 
photometry of the younger, more luminous main-sequence and blue loop stars. 
The images are available online at the HST Data Archive under the listing
for Program ID\# 6865.

\begin{figure*}
\centerline{\hbox{\psfig{figure=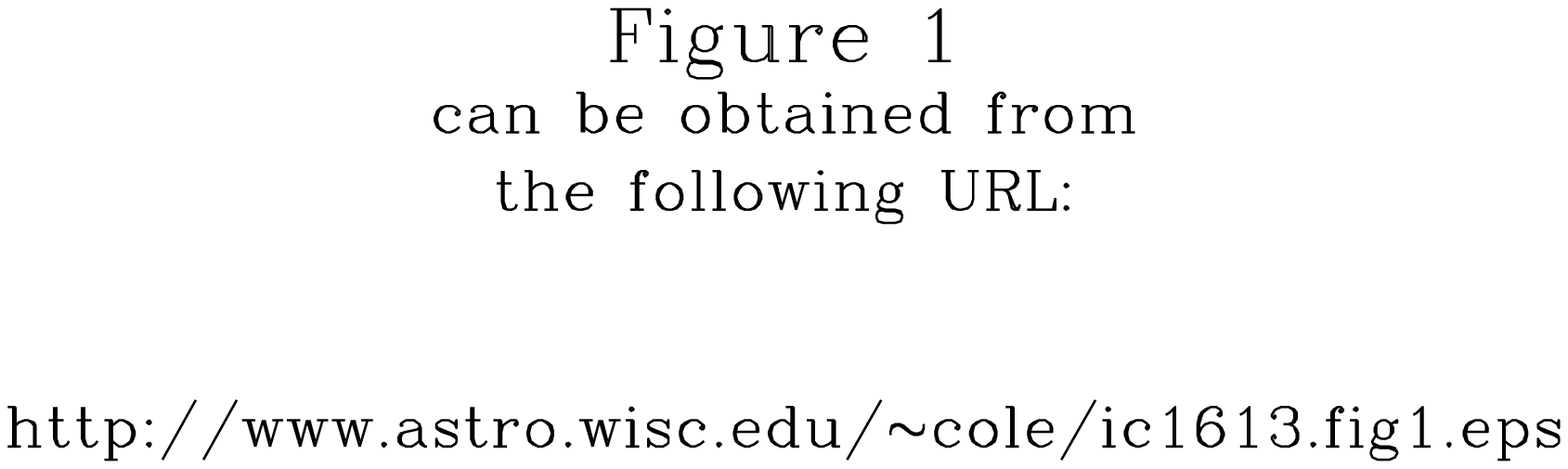,height=3.0in}}}
\caption{WFPC2 F555W image of the center of \ic.  North and East 
are as marked; the field of view is roughly 3$\farcm$5 across the widest
corner-to-corner span.  The low stellar density of \ic\ is highlighted
by the large number of background galaxies in the image.  \label{wfpc}}
\end{figure*}

All images were reduced using the standard pipeline
at the Space Telescope Science Institute (Holtzman {\it et al.} 1995a), 
and combined using a 
cosmic-ray cleaning algorithm.  The combined F555W image is shown
in Figure \ref{wfpc}.  The field of view was oriented
to place the center of \ic\ in the WF3 chip; most of the central
\hi\ hole (Skillman 1987; Lake \& Skillman 1989) is contained 
within the five square-arcminute field.  Additionally, the field
was chosen to avoid \hii\ regions (Hodge, Lee \& Gurwell 1990) 
in order to focus attention on the older main-sequence stars that
cannot be photometered from the ground due to crowding.
The low stellar and gas
density of \ic\ are clearly evident by the clarity with which 
background galaxies can be seen.  

\subsection{Photometry}

We performed profile-fitting photometry on our reduced, combined
images using the DoPHOT routines and techniques described in Schechter, Mateo
and Saha (1993); 
the code has been modified to handle the complexities of the WFPC2 point
spread function (Saha {\it et al.} 1994).  
Using the DoPHOT star-matching routines,
we identified 47,263 stars in both the F555W and F814W frames; the shallower
F439W exposures permitted the cross-identification of just 7,922 stars in
the F439W and F555W images.  Our data have limiting magnitudes for photometry
of V $\sim$ 26.5, I $\sim$ 25.5, and B $\sim$ 25.

\begin{figure*}
\centerline{\hbox{\psfig{figure=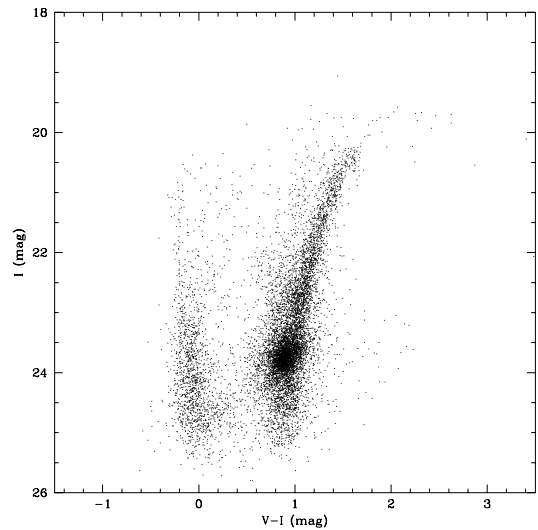,height=3.0in}}}
\caption{WFPC2 I, V$-$I color-magnitude diagram, including only
stars for which DoPHOT achieved the best possible photometry.  9800 stars
are shown.  The most striking features of this CMD are the prominent
red giant branch and the upper main-sequence.  \label{vicmd}}
\end{figure*}

\begin{figure*}
\centerline{\hbox{\psfig{figure=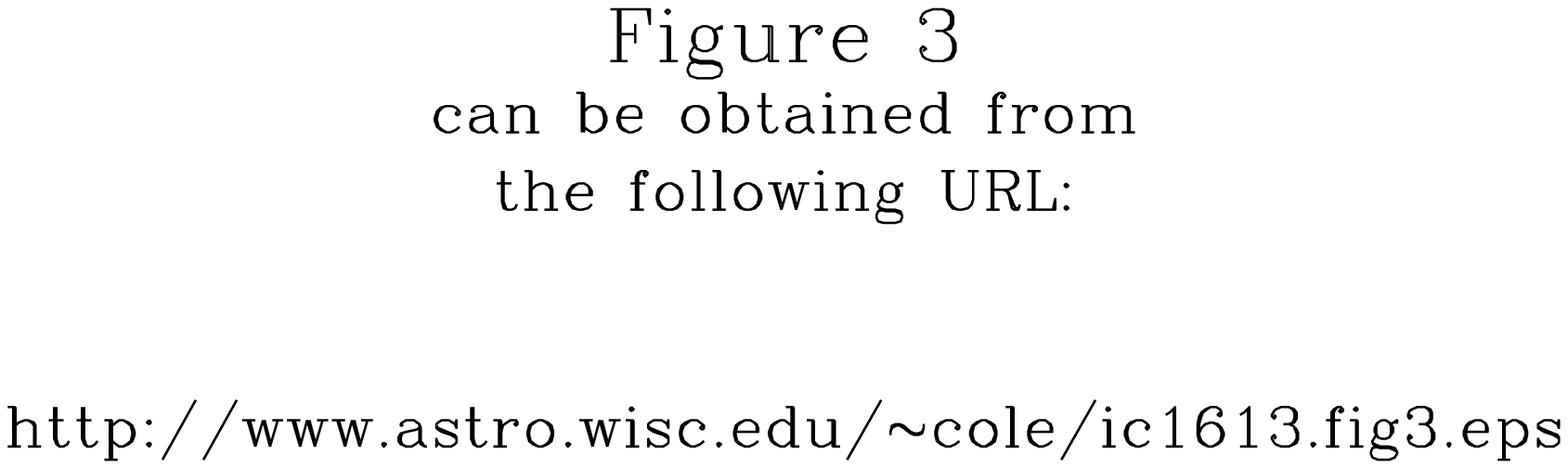,height=3.0in}}}
\caption{The distributions of magnitude vs. photometric error
for our WFPC2 BVI photometry.  The mean errors at 24th magnitude are 
$\sigma_V$ = 0.03, $\sigma_I$ = 0.04, $\sigma_B$ = 0.03.
\label{errfig}}
\end{figure*}

We transformed our photometry to the Johnson-Cousins BVI system
by making aperture corrections to the 0$\farcs$5 aperture recommended
by \cite{hol95b}, adopting the STMAG system zeropoints therein,
and converting the resulting values to the BVI system using the 
iterative procedure described in \cite{hol95b}, with coefficients from 
their Table 10.  Figure \ref{vicmd} contains our final WFPC2 (I, V$-$I)
color-magnitude diagram for \ic.  Typical 1-$\sigma$ errorbars
at 24th magnitude are $\sigma _{\mathrm V}$ = 0.03, $\sigma _{\mathrm I}$ = 0.04.
The distributions of magnitude vs. error are shown in Figure \ref{errfig}. 

Along with magnitudes and standard errors,
DoPHOT returns for each object a ``quality'' parameter, which allows the 
astronomer to discriminate true point sources from unresolved blends, 
background galaxies, and stars contaminated by cosmetic CCD blemishes.
Because of the steep main-sequence luminosity function, the blending 
criterion most strictly winnows the photometric sample at faint magnitudes.
For this paper, we have retained in our analysis only those stars with
the highest quality photometry (``Type 1'').  The remainder of this
paper considers only
the 9,810 stars with superior F555W and F814W photometry, and the 1,639
stars with superior F439W photometry.  Because of the small numbers of
luminous, blue stars in our chosen field, the F439W photometry has been
used solely to check our adopted reddening values for internal consistency.

The smoothly rising error curves in Figure \ref{errfig}
indicate that color-magnitude diagrams will not be
systematically distorted by the effects of crowding (Dohm-Palmer {\it et al.}
1997a, Han {\it et al.} 1997).  From experience with WFPC2 photometry,
we estimate our 90\% and 50\% completeness levels in (V,I) fall at 
(23.8, 23.8), and (25.6, 25.2), respectively.  Completeness issues are
most problematic for the lower main-sequence and for the giant branch
below the level of the red clump.  Completeness is high over most of the
magnitude range in Figures \ref{vicmd} and \ref{errfig}, decreasing smoothly
towards fainter magnitudes; because most of our analysis deals with stars
brighter than 24th magnitude, incompleteness is not expected to dramatically
affect our results.

\subsection{Observations with WIYN}

\begin{figure*}
\centerline{\hbox{\psfig{figure=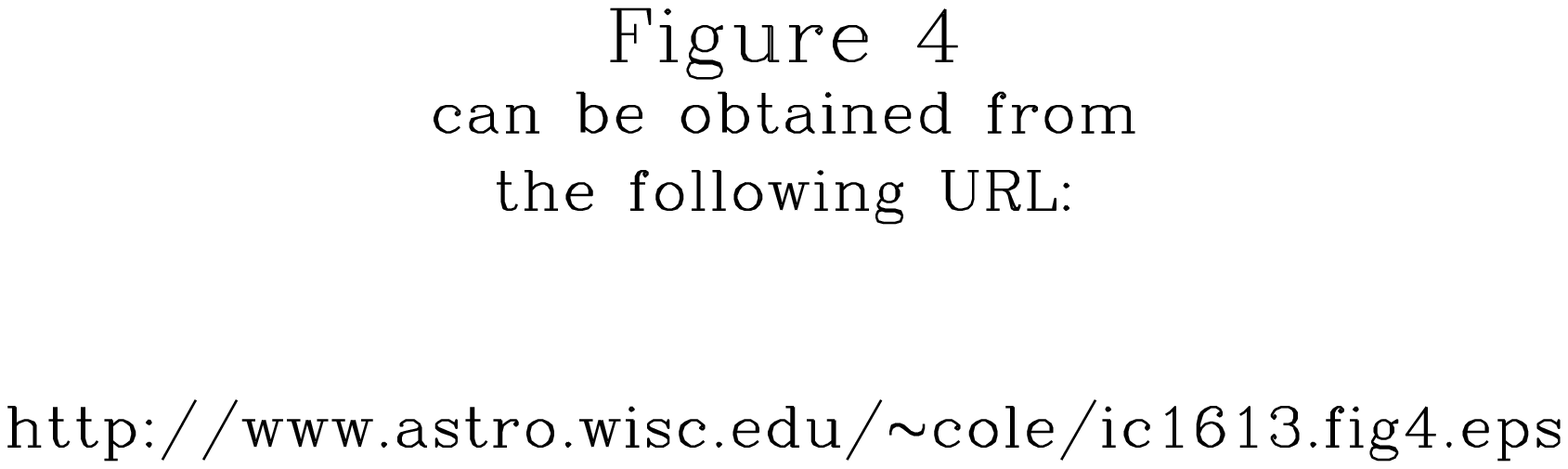,height=3.0in}}}
\caption{900-second, V band, WIYN image of the main body of \ic.
North is up, East is to the left, and the WFPC2 field of view is overlain.
The seeing is 0$\farcs$6, causing the background galaxies to stand out
clearly through the stars of \ic.  \label{wiyn}}
\end{figure*}

To broaden our areal coverage of \ic, we obtained V and I
images centered on the WFPC2 position with the 3.5-meter WIYN
telescope at Kitt Peak,
on 1--3 October, 1997.  The best data, comprising 2$\times$600 seconds
in I and 900 seconds in V, were 
taken under photometric conditions with 0$\farcs$6 seeing.  Additional
exposures with $\sim$1$\arcsec$ seeing totalling 1500 sec in each filter
were obtained.  The broad (6$\farcm$8 square--- $\approx$9 times the
area of the WFPC2 field) field of view and fine image 
quality of the WIYN imager 
allow us to extend our CMD analysis to luminous, short-lived phases of
stellar evolution which are typically under-represented in the WFPC2 field.  
A WIYN image of \ic, with the WFPC2 field overlain for comparison, is
shown in Figure \ref{wiyn}.

\begin{figure*}
\centerline{\hbox{\psfig{figure=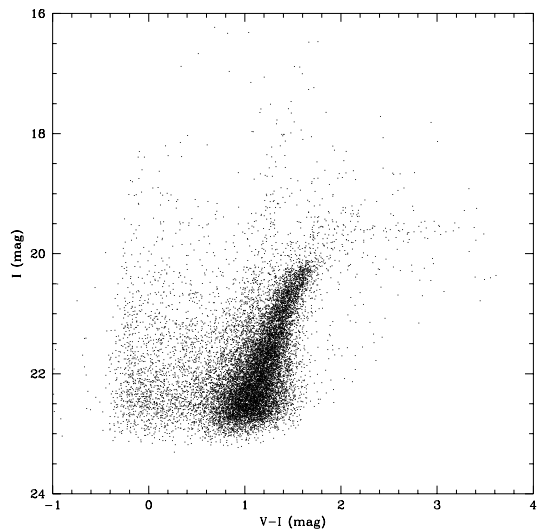,height=3.0in}}}
\caption{WIYN I, V$-$I color-magnitude diagram, including only
stars for which DoPHOT achieved the best possible photometry.  14678
stars are shown.  Image crowding has blended the red giant branch
with blue loop stars and red supergiants, causing an artificially large
color spread.  The luminous red supergiants and extremely red tail of
the asymptotic giant branch are more fully sampled by the 
6$\farcm$8 $\times$ 6$\farcm$8 field of view.   \label{wiyncmd}}
\end{figure*}

The WIYN images were overscan corrected, bias-subtracted, and flat-fielded
within IRAF's {\tt ccdproc} routines\footnote{
IRAF is distributed by the National Optical Astronomy Observatories,
    which are operated by the Association of Universities for Research
    in Astronomy, Inc., under cooperative agreement with the National
    Science Foundation.}.  We performed photometry using DoPHOT,
as per the WFPC2 images, obtaining ``Type 1'' photometry for 14,678 VI pairs.
We also observed standard fields from Landolt (1992) to reduce the data 
to a standard system for comparison to the WFPC2 data.  The WIYN color-magnitude
diagram is shown in Figure \ref{wiyncmd}.  Comparison to Fig. \ref{vicmd} shows
good agreement regarding the magnitude of the tip of the red giant branch and
the colors of the red giant branch and main-sequence.  Comparison 
of individual stars on the red giant branch
finds I$_{\mathrm WIYN}$ = I$_{\mathrm WFPC2}$ $-$0.04 
$\pm$0.03, (V$-$I)$_{\mathrm WIYN}$ =
(V$-$I)$_{\mathrm WFPC2}$ + 0.04 $\pm$0.04.

\section{Color-Magnitude Diagrams}
\label{cmd}

Figures \ref{vicmd} and \ref{wiyncmd} show two main branches,
and several smaller features, each of which contains stars with a 
range of ages and metallicites:

\begin{enumerate}

\item The red giant branch (RGB), between V$-$I = 0.7--1.7 and extending
upwards to I = 20.3; the strongest feature of the CMD, contains
stars from 500 Myr to $\gtrsim$10 Gyr among its various substructures.
The location of the RGB tip agrees well with the adopted distance to \ic.

\item The blue plume, blueward of V$-$I = 0.2, consisting primarily
of main-sequence stars younger than $\approx$1 Gyr. 
The majority of blue plume stars in Figure \ref{vicmd}
are expected to be of spectral class B; 
Figure \ref{wiyncmd} includes some probable
O stars, indicators of very recent ($\lesssim$ 10 Myr) star formation.

\item The red clump (RC), centered at I = 23.76, V$-$I = 0.9.  This feature
contains core-helium burning stars with ages between $\approx$ 1--10 Gyr, and 
accounts for a plurality of the stars in Figure \ref{vicmd}.

\item The blue loop (BL) stars, extending vertically from the RC to 
I $\approx$ 22.5 before turning diagonally blueward to become indistinguishable
from the main sequence at I $\approx$ 21.  These are the younger counterparts
to the RC stars.  Crowding in Figure \ref{wiyncmd} renders them 
indistinguishable from the RGB.

\item The asymptotic giant branch (AGB), extending redward from the tip
of the RGB, is weakly present in Figure \ref{vicmd}, but more strongly apparent
in the WIYN CMD.  AGB stars indicate an intermediate-age population
(2--6 Gyr).  A horizontal AGB indicates metallicities of $\sim$ 1/10 solar;
additionally, a more metal-poor component lies along an upward extension 
of the RGB in Figure \ref{wiyncmd}.

\item The red supergiants (RSG), extending nearly vertically to I = 16.5 
at V$-$I $\approx$ 1.3 in Figure \ref{wiyncmd}.  Fewer than 16 of these are
expected to be foreground dwarfs (Bahcall \& Soneira 1980); their possible
presence does not compromise the reality of the \ic\ RSG sequence.
These young stars are
on their way to an `onion-skin' nuclear burning structure and are younger 
than a few 10$^7$ years.  Such young stars should be representative of the
most metal-rich population in \ic.

\item The horizontal branch, visible at the base of the RC as a blueward
extension at I = 24.3, and possibly as a red extension from the main-sequence
at I = 24.7.  These are the most metal-poor, oldest (10--15 Gyr) core-helium burning 
stars in \ic, tracers of a population whose main-sequence turnoff would
be expected to lie at I $\approx$ 28.  The gap in the horizontal
branch between 0.3 $\leq$ V$-$I $\leq$ 0.6 is consistent with the 
location of the RR Lyrae instability strip in the CMD.

\end{enumerate}

\begin{figure*}
\centerline{\hbox{\psfig{figure=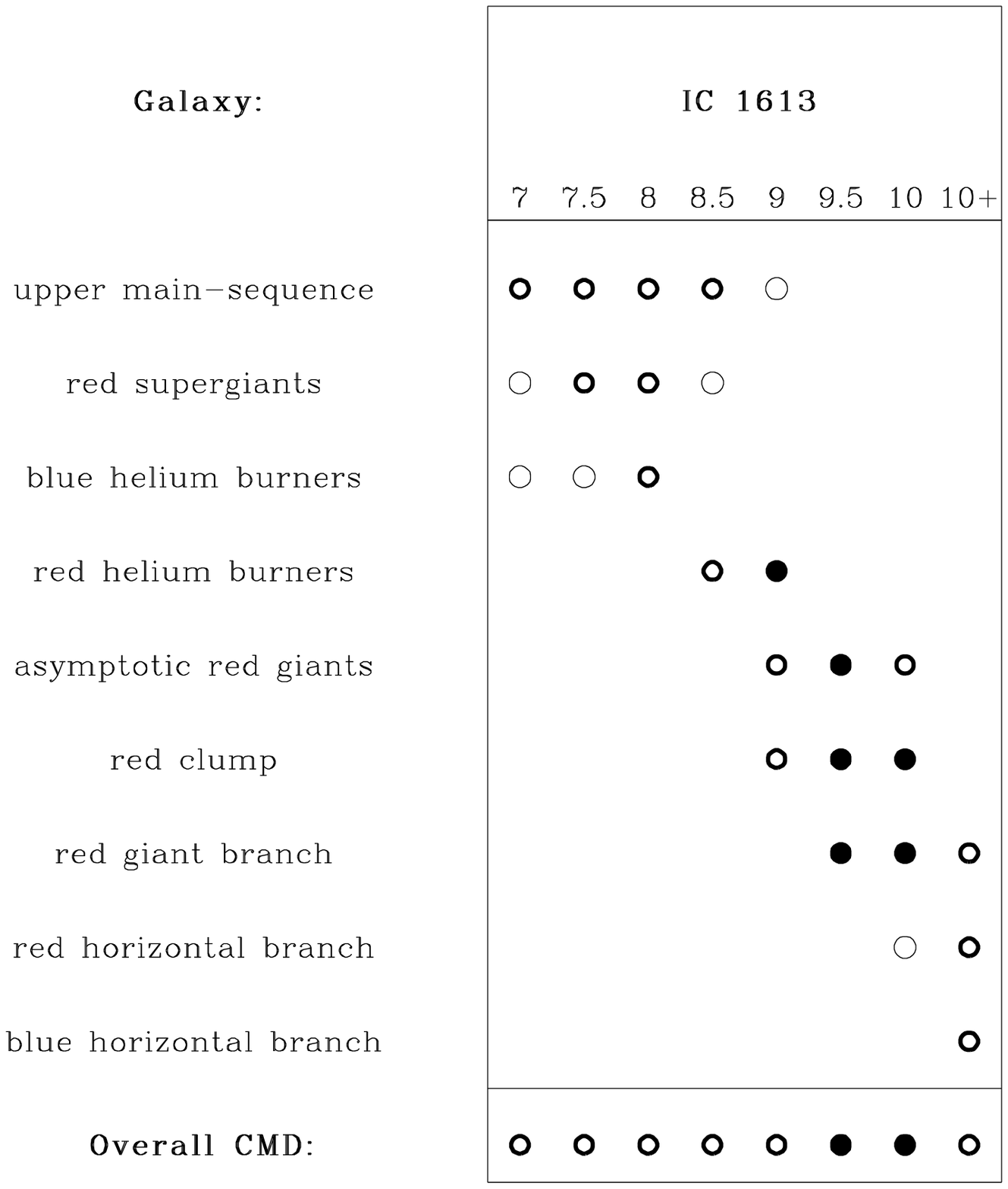,height=5.0in}}}
\caption{Reference card for the stellar populations at the 
center of \ic\ as derived from the morphology of WFPC2 and WIYN
color-magnitude diagrams.  CMD features are associated with populations
of various ages.  The relative strength of each population is given by
the shaded circles in the appropriate logarithmic age bins.  A general
picture of approximately continuous star formation emerges, although 
our ability to resolve ``bursty'' behavior becomes progressively poorer
with age.   \label{score}}
\end{figure*}

The WFPC2 and WIYN color-magnitude diagrams
contain stars as young as the dynamical timescale for protostellar
collapse and potentially as old as the Galactic globular clusters.
The salient features of the CMDs are summarized in
Figure \ref{score}.  The dividing point between the blue and 
red helium burners (HeB) has been defined here to be the Cepheid instability 
strip, by analogy to the blue and red horizontal branches.
Each feature is marked according 
to the approximate logarithmic age expected for the
given evolutionary phase; the prominence of the marker
symbols provides a rough guide to the strength of the
components of the CMD.  Taken together, the CMDs 
provide an overall picture of the star-formation history
of \ic: it appears to have formed stars over wide ranges
of age, perhaps continuously; most of the star-formation
may have taken place a few to ten billion years ago.
In subsequent sections of this paper, we examine the
implications of CMD morphology for the star-formation history
of \ic\ in more detail.

We take the distance to \ic, from Lee {\it et al.} (1993), to be 715 kpc,
corresponding to a 
distance modulus (m-M)$_0$ = 24.27.  We correct for a small
amount of interstellar reddening, E$_{\mathrm B-V}$ = 0.03 (Mateo 1998).
We derive a value E$_{\mathrm B-V}$ = 0.02 $\pm$0.02 from a comparison
of the B$-$V and V$-$I colors of the upper main-sequence between 
24 $\geq$ V $\geq$ 23, using the intrinsic color of the zero age 
main-sequence as a template.  We adopt the Galactic reddening law
of Cardelli, Clayton \& Mathis (1989), using R$_V$ = 3.1, to derive
extinction parameters A$_{\mathrm V}$ = 0.093, A$_{\mathrm I}$ = 0.045,
and E$_{\mathrm V-I}$ = 0.048.

\section{The Stellar Populations of \ic}
\label{pop}

The mean color of the red giant branch, in a simple stellar population,
can be used to estimate the metallicity of its stars.
Da Costa \& Armandroff (1990) give a 
calibration for globular clusters' metallicities using the mean,
dereddened V$-$I color of their giant branches at an absolute magnitude
M$_{\mathrm I}$ = $-$3.  At the distance and reddening of \ic, this
translates to I = 21.25.  From the WFPC2 CMD, we use 47 stars to determine
$\bar{V-I}_0$ = 1.29 $\pm$ 0.07;
the much larger sample of 255 stars in the
WIYN field yields
$\bar{V-I}_0$ = 1.28 $\pm$ 0.10.
The Da Costa \&
Armandroff (1990) calibration gives [Fe/H]$_{\mathrm WFPC2}$ = $-$1.38
$\pm$ 0.31, and [Fe/H]$_{\mathrm WIYN}$ = $-$1.43 $\pm$ 0.45.  These
abundance spreads are not solely due to photometric errors, which 
limit the resolution in [Fe/H] to 0.17 dex for WIYN, and 0.13 dex
for WFPC2; an intrinsic spread in color is present due to ranges in
age and/or metallicity.
Our estimate of [Fe/H] $\approx$ $-$1.4
is in excellent agreement with the previous assessment
[Fe/H]$_{\mathrm RGB}$ = $-$1.3 $\pm$0.8 (Freedman 1988a).
Although \ic\ is certainly {\it not} a simple, single age stellar population,
this calculation still provides a useful first estimate of the stellar
metallicities (e.g., Da Costa 1998).

A real abundance variation is almost certainly
present in \ic.  However, 
the presence of intermediate-age (5--10 Gyr) stars, strongly suggested
by the prominent red clump at I $\approx$ 23.8, complicates the estimation
of metallicity from RGB color.  In stellar populations that are significantly
younger than the globular clusters, the giant branches lie to the blue of
the Da Costa \& Armandroff (1990) calibration sequence, causing the 
$\bar{V-I}_0$
method to underestimate the
metallicity by up to $\approx$ 0.3
dex (see, e.g., Da Costa 1998).  This is the
infamous age-metallicity degeneracy;
the inferred metallicity range among \ic's red giants may be, in part, due
to an age range.  If age and metallicity are correlated (e.g., with chemical
enrichment occuring over time), than it is possible to underestimate the 
range of metallicities among giant branch stars.  Independent abundance
information, together with detailed modelling of the star-formation
history, is needed in order to lift the age-metallicity degeneracy.

To test our photometry, we rederived the distance to \ic\ using
the tip of the RGB in our WFPC2 CMD.  We followed the metallicity
calibration given by Lee {\it et al.} (1993), and applied a 
Sobel filter (Sakai, Madore \& Freedman 1996) to the I-band
luminosity function.  This fixed the tip of the RGB at I = 20.28 $\pm$0.09,
which yields a distance
modulus (m$-$M)$_0$ = 24.29 $\pm$0.12 for E$_{\mathrm B-V}$ = 0.03 $\pm$0.02
and [Fe/H] = $-$1.38 $\pm$0.31.  This is in excellent agreement with the
adopted value of (m$-$M)$_0$ = 24.27 $\pm$0.10.

\subsection{Recent Star-Formation History}

The young stellar populations of \ic\ have been studied by Sandage \& Katem
(1976), Hodge (1978), Freedman (1988a,b), Hodge {\it et al.} (1991), and
Georgiev {\it et al.} (1999).  The picture that emerges from these 
studies is of a galaxy in which the central regions have undergone
continuous star formation over the past 1-300 Myr, with the youngest
stars clustered into small associations and possible clusters (Hodge
1978).  While the regions of recent star formation are more centrally
concentrated than the underlying old population, the very youngest
stars seem to avoid the dead center of \ic\ (Hodge {\it et al.} 1991).
This partially accounts for the lack of blue stars brighter than
M$_{\mathrm I}$ $\approx$ $-$3.5 in the WFPC2 field of view. 

\begin{figure*}
\centerline{\hbox{\psfig{figure=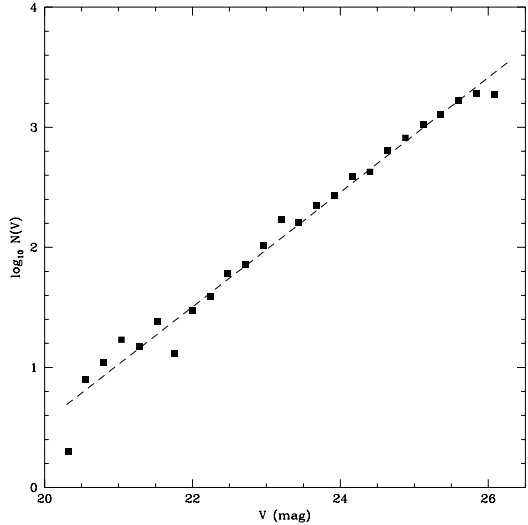,height=3.0in}}}
\caption{Main-sequence V-band WFPC2 luminosity 
function for all photometry.
Incompleteness becomes a serious problem at V $\approx$ 24.5.  The least-squares
fit slope is $-$0.48 $\pm$0.09; this includes stars as old as $\approx$
350 Myr. Normalized for a Salpeter IMF this implies a mean SFR over that time
of 3.5 $\times$ 10$^{-4}$ M$_{\sun}$ yr$^{-1}$ across the 0.22 kpc$^2$ WFPC
field of view.   \label{vmslf}}
\end{figure*}

Freedman (1988a) found that the main-sequence
luminosity function brighter than
M$_{\mathrm V}$ = 21 could be fit with a simple power law, similar to
other Local Group dwarf irregulars, with N(m) $\propto$ m$_V^{0.58\pm0.07}$.
Freedman attributes a flattening of the luminosity function
slope at V $\approx$ 21 to crowding in her data.  The WFPC2 V-band luminosity 
function is plotted in Figure \ref{vmslf}.  We find 
N(m) $\propto$ m$_V^{0.48\pm0.09}$,
marginally consistent
with the steeper slope from Freedman (1988a)\footnote{The luminosity function
of \ic\ is thus nearly identical to that of the Sextans A dwarf irregular, 
for which Dohm-Palmer {\it et al.} (1997a) found N(m) $\propto$ m$_V^{0.44}$.}.
Statistical noise begins
to dominate the WFPC2 main-sequence luminosity function brighter than 
V $\approx$ 21.5; incompleteness due to crowding becomes problematic below
V $\approx$ 24.5.  Within these limits, which roughly bracket the range
of the B stars (3--20 M$_{\sun}$; ages between 10--350 Myr), the constant
slope of the logarithmic luminosity function suggests
an approximately constant star-formation rate.
We integrate a Salpeter IMF from 0.1--120 M$_{\sun}$ to normalize the
upper main-sequence
luminosity function, and find a total star-formation rate of 
3.5 $\times$ 10$^{-4}$ M$_{\sun}$ yr$^{-1}$ across the WFPC2 field.
At the distance of \ic, this corresponds to 1.6 $\times$ 10$^{-3}$
M$_{\sun}$ yr$^{-1}$ kpc$^{-2}$.

For comparison, this is slightly smaller than the current star-formation
rate in the brightest \hii\ region (S10, a.k.a.\ complex 67), although
S10 fills just one-tenth the area of the WFPC2 field and therefore sustains
a much higher {\it intensity} of star-formation.  This behavior is common
to dwarf irregular galaxies; for example, Dohm-Palmer {\it et al.} (1997b)
found that star formation in Sextans A has occurred in localized
regions $\approx$ 10$^2$--10$^3$ parsecs in size, and that these regions
last for 100--200 Myr at a time.

Davidson \& Kinman (1982) and Kingsburgh \& Barlow (1995) have studied
the Wolf-Rayet star DR 1 and its surrounding \hii\ region; this object,
presumably a few million years old, has an abundance between 
Z $\approx$ 0.004 ([Fe/H] = $-$0.7, Kingsburgh \& Barlow 1995)
and [O/H] = $-$1 (Davidson \& Kinman 1982).
These are in reasonable agreement with
the \hii\ region nebular abundances reported by 
Skillman, Kennicutt, \& Hodge (1989) of [O/H] $\approx$ $-$1.0.
Recent B star spectra (Kudritzki 1999) provide further support
for the case that \ic\ is currently forming stars with nearly
the same metallicity as the Small Magellanic Cloud (Hodge 1989).
The region of our present study was chosen to avoid known
\hii\ regions, and it is of interest to determine the highest metallicity
attained during recent (but not current) star formation.

We can estimate the metallicity of the youngest stellar population
in the WFPC2 field of view
from the color extent of the blue loop stars--- massive, core-helium
burning stars--- and the slope of the red supergiant sequence in
the color-magnitude diagram.  The WIYN CMD shows a well-populated
RSG sequence, rising to I $\approx$ 16.5; the spread in color of 
this sequence can be due to age as well as metallicity.  From a 
comparison to theoretical isochrones (Girardi {\it et al.} 1999),
the majority of the RSGs cannot
be as metal-rich as Z = 0.004 ([Fe/H] $\approx$ $-$0.7); the Z = 0.001
([Fe/H] $\approx$ $-$1.3) isochrones provide a better fit to the data.

\begin{figure*}
\centerline{\hbox{\psfig{figure=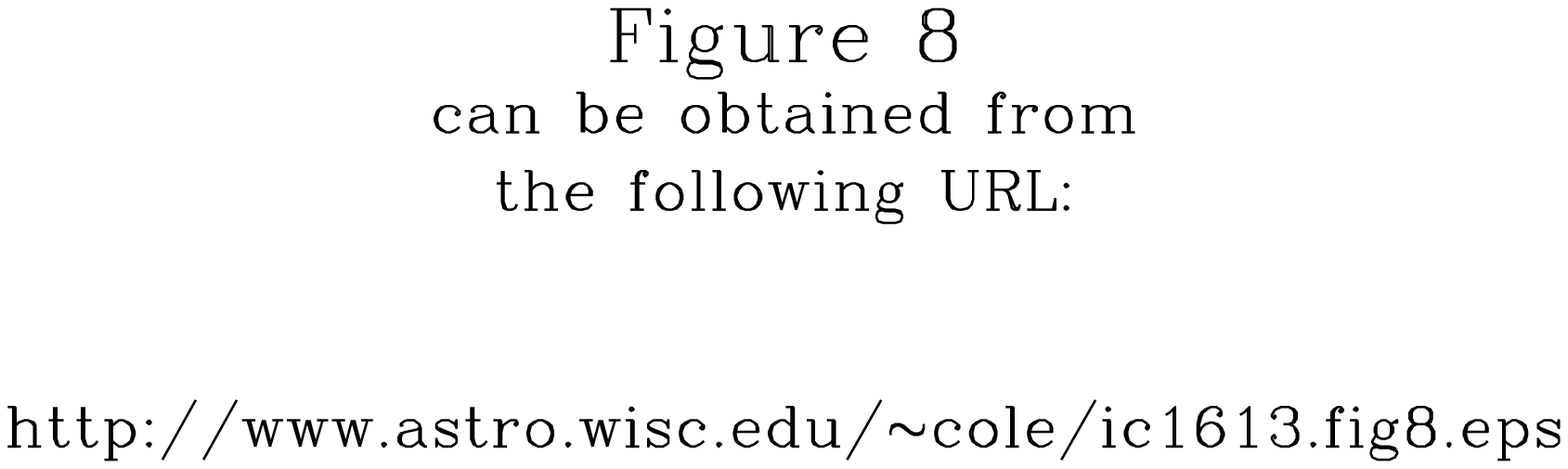,height=3.0in}}}
\caption{Synthetic CMDs created assuming a Salpeter IMF and
constant star formation rate from 250--1000 Myr.  The Hess diagram
for \ic\ is overlain.  Left panel: model Z = 0.001.  Right panel: model
Z = 0.004.  The red giant branch, older than 1000 Myr, has not been
modelled here so that the locus of young giants and supergiants can be
seen more clearly.   \label{blfig}}
\end{figure*}

While the RSG sequence above I $\approx$ 20 is underpopulated in the
WFPC2 CMD, the blue loop stars are prominent.  We used a Monte Carlo
code to compare the isochrones of Girardi {\it et al.} (1999) to this
portion of the \ic\ Hess diagram.\footnote{a contour plot of the number of stars
per unit magnitude per unit color in the CMD--- see Hess (1924);
Trumpler \& Weaver (1953)}  Figure \ref{blfig} shows the results of
this simulation.  We have sampled from the isochrone set assuming a 
constant star formation rate and Salpeter (1955) initial mass function
for two metallicities, Z = 0.001 (left panel), and Z = 0.004 (right
panel).  In each panel of Figure \ref{blfig} the star-formation 
rate was constant from 0.25 -- 1 Gyr. The
Hess diagram of the WFPC2 data is overplotted against the simulations,
with isopleths from 2 -- 256 decimag$^{-2}$; each successive 
isopleth shows twice the CMD density of the previous one.

Figure \ref{blfig} affords two main points of comparison between
models and data.  Firstly, the red HeB (blue loop) stars between I $\approx$
23.5 and 22, with V$-$I $\lesssim$ 1; secondly, the base of the RSG
plume with I $\lesssim$ 22 and V$-$I $\approx$ 1.  Dohm-Palmer
{\it et al.} (1997b) presented the first unambiguous detection of the
blue loop stars in a metal-poor system and showed how these stars
can be used to infer the star formation history of a galaxy over
the past 1 Gyr.  The clearest
inference from Fig. \ref{blfig} flows from the termination of the 
blue loop sequence at I $\approx$ 22; this
morphology indicates a higher past star-formation rate that declined
at some point; the age of decline is metallicity dependent: 300 Myr
for Z = 0.004, 400 Myr for Z = 0.001.  The extent of the blue loop
sequence in the models is clearly metallicity dependent; we see from
Fig. \ref{blfig} that the Z = 0.001 models are bluer than the 
data, while the Z = 0.004 models are too red.  The same behavior
can be seen with respect to the RSG sequence.  An intermediate
metallicity, Z $\approx$ 0.002, is therefore predicted for the 
0.1 -- 1 Gyr old stellar populations of \ic.  This value is in 
excellent agreement with the nebular abundances of Skillman {\it 
et al.} (1989) and Davidson \& Kinman (1982).

We have not directly compared the \ic\ Hess diagram to the WIYN
CMD; these data do not reach beyond 100 Myr on the main-sequence,
and are thus unmodelled by the Girardi {\it et al.} (1999) isochrone
set.  As in the WFPC2 CMD, the Z = 0.004 models are too red when
compared to the RSG plume.  Even the Z = 0.001 model blue loops
do not reach far to the blue as the blue loops in the WIYN CMD
above I $\approx$ 21.5.  If the models, and the photometry, 
can be trusted on this point, this would be indicative of 
a metallicity well below Z = 0.001 for the youngest generation of
stars in \ic.  However, comparison of the WIYN and WFPC2 CMDs 
shows that crowding errors affect even the brighter stars in 
ground-based images (see, e.g., Tolstoy {\it et al.} 1998).
Additionally, it has been shown that the color extent
of blue loops in models of massive stars is quite sensitive to the adopted
input physics (e.g., Chiosi 1998; Langer 1998); effects such as the mass-loss 
prescription, semiconvection,  and  rotation (e.g., Langer \& Maeder 1995)
can lead to a wide variety of blue loop morphologies 
at a given age and metallicity. 

\subsection{Intermediate-Age and Old Stellar Populations}

Baade (1963) and Sandage (1971a) noted the elliptical sheet
of red stars within which the main body of \ic\ is embedded;
drawing on Baade's (1944) division of the stellar populations
of the more luminous Local Group galaxies into Populations I and
II, this sheet has been identified with the Pop II element of
\ic.  Freedman's (1988a) photometry resolved this sheet into 
red giant and aymptotic giant branches, to a level of V $\approx$ 23,
confirming the presence of stellar populations aged at least $\approx$1
Gyr, and perhaps as old as the most ancient Galactic stars, 10--15 Gyr old.

Our WIYN color-magnitude diagram extends Freedman's (1988a) results;
the strongest feature of the WIYN CMD (Fig. \ref{wiyncmd}) is the classic
Pop II red giant branch.  We detect red giants to a limiting magnitude
of I $\approx$ 23 (V $\approx$ 24) across the whole of the 6$\farcm$8 $\times$
6$\farcm$8 field of view.  Although the center of \ic\ has not seen
significant recent star formation and is the site of a hole in the \hi\ 
distribution (Lake \& Skillman 1989), the stars of our WIYN image
(Fig. \ref{wiyn}) show no tendency to avoid this area.  This is in agreement
with the results of Hodge {\it et al.} (1991), who noted the differing 
behaviors of the bright blue stars and the faint red stars
in this regard.

Because of the near-degeneracy of the red giant branch V$-$I color 
to age beyond 2 Gyr, it has not previously
been possible to infer the age mixture
of the Baade's sheet population in \ic\ from CMD morphology (Freedman
1988a); to resolve this issue requires photometry further down the
main-sequence.
Saha {\it et al.} (1992) identified 15 RR Lyrae-type variable
stars in \ic; these betray the presence of a genuinely ancient
stellar population in this galaxy.  The observed variation in 
RR Lyrae production effciency among Galactic globular clusters
is evidence of an unknown (``second'') parameter which impacts 
the horizontal branch morphology of ancient stellar populations
(van den Bergh 1967; Lee {\it et al.} 1994). 
This effect limits the power RR Lyrae 
number counts to constrain the precise contribution of ancient
stars to the \ic\ RGB; a lower limit of 15\% was placed on 
the fraction of red giants older than $\approx$10 Gyr.

\subsubsection{The Horizontal Branch}

The presence of a red horizontal branch (RHB) population in the WFPC2
VI color-magnitude diagram can be discerned, in the color range
V$-$I $\approx$ 0.6--0.8, at I $\approx$ 24.2--24.3.  This is consistent
with the mean $g$ magnitude (Thuan \& Gunn 1976) of 24.9 found by
Saha {\it et al.} (1992) for the \ic\ RR Lyraes.  The Pop II stars
of the central 500 pc of \ic\ thus seem similar
to the those in the relatively distant field studied by Saha {\it et al.}
(1992).  The theoretical RHBs of recent isochrone
sets (e.g., Girardi {\it et al.} 1999) predict a magnitude of I $\approx$
24.35, for a metal abundance of 1/19 solar (Z = 0.001, corresponding
to [Fe/H] = $-$1.3) at the distance of \ic.  The $-$0.15 mag offset 
between the observed RHB and model isochrones can
be accounted for by a difference in [Fe/H] of $\approx$ $-$0.5 to $-$0.7 dex
(Caloi, D'Antona \& Mazzitelli 1997), suggesting a metallicity for the
RHB population of [Fe/H] $\approx$ $-$1.8 to $-$2.0.

The red spur which extends from the main-sequence to V$-$I $\approx$
0.3 at I $\approx$ 24.7--24.8 cannot be definitively assigned to a
specific evolutionary state without further analysis.  At least two
different classes of stellar population can create such a feature in 
the color-magnitude diagram. The red spur could be post-main-sequence-turnoff
stars aged 900 to 1100 million years with Z = 0.001;
alternatively, its color and magnitude are consistent with the blue horizontal
branch (BHB) of a 10--15 billion year old population of very low metallicity
(Z $\approx$ 0.0004).

The BHB interpretation shows consistency with previously known properties
of \ic: its old average age, the inferred metallicity of the red horizontal
branch stars, and the detection of RR Lyrae variables by Saha {\it et al.}
(1992).  It is nigh impossible to precisely predict the expected number of BHB stars
from \ic's RGB luminosity function, because the RGB is likely contaminated
with intermediate-age stars, and because the second parameter effect decouples
horizontal branch morphology from the known physical properties of ancient 
stellar populations.  Nevertheless, we performed a simple analysis to estimate the
relative star-formation rates required to fully populate the red spur with
BHB stars.  We used a Monte Carlo technique to draw stellar populations from
the Girardi {\it et al.} (1999) isochrones according to the initial mass
function of Salpeter (1955) from 0.7 -- 5 M$_{\sun}$, to reproduce
the relative numbers of stars seen in the red spur and on the main-sequence at
the same I magnitude. From this we infer that the 
star-formation rate 15 Gyr ago was some 2--200
times larger than the recent (t $\lesssim$ 0.9 Gyr) star-formation
rate.  The limiting uncertainty in this calculation is the second parameter
effect (see Saha {\it et al.} 1992).  The main-sequence turnoff associated
with this dominant, ancient population would be located at I $\approx$
27.5--28, requiring an 8-meter class telescope to photometer reliably.

We used the same analysis technique to evaluate the hypothesis that the 
red spur at I $\approx$ 24.7 is the subgiant branch produced by
a burst of star formation $\approx$ 1 billion years ago.  We divided the
stars in this region of the CMD (25 $>$ I $>$ 24.3) into two color bins:
a ``main-sequence'' bin from $-$0.4 $<$ V$-$I $<$ 0.1, and a ``red spur''
bin from 0.1 $<$ V$-$I $<$ 0.5.  The number ratio of main-sequence stars
to spur stars is
2.08, which can be reproduced with models in which the star-formation
rate from 900 to 1100 million years ago was a factor of 30 higher than
before or since.  

This large increase in star-formation rate
would have dramatic consequences for the color-magnitude diagram of \ic,
as well as for the galaxy itself.  A sharp discontinuity in the 
WFPC2 main-sequence
luminosity function would be present at I $\approx$ 25.1, corresponding 
to the main-sequence turnoff of the putative 1 Gyr burst. 
Past experience with WFPC2 photometry (e.g., Dohm-Palmer {\it et al.}
1997a, Gallagher {\it et al.} 1998) suggests that our photometry is
$\approx$50\% complete at this level.  Using all of the WFPC2 photometry,
we find rising power-law luminosity function down to I $\approx$ 26 
with a sharp decline due to incompleteness below that level.

Although increasing photometric errors prevent the lower main-sequence
luminosity function from {\it absolutely} ruling out a
factor of 30 increase in star-formation rate a billion years ago,
the stars of the red clump and
giant branch are not compromised by these problems.  A large 1 billion
year-old population will produce a very distinctive ``vertical red clump'',
or pseudo-clump, concentrated in color between 0.7 $\lesssim$ V$-$I $\lesssim$
0.9 and stretching from M$_I$ $\approx$ 0 to $-$1.2 (Caputo, Castellani \& 
Degl'Innocenti 1995; Girardi {\it et al.} 1998). 
The model which best reproduces the
ratio of stars in the red spur to stars on the main-sequence predicts
a ratio of pseudo-clump stars to brighter RGB stars of 6.8, which is  
inconsistent with our observed ratio of 1.5.  While \ic\ shows a pseudo-clump,
its extension to I $\approx$ 22 (M$_I$ $\approx$ $-$2.25) suggests a period
of more or less continuous star formation from 400--1200 Myr ago.

A further inconsistency with the 
1 Gyr burst scenario for the red spur is the relative lack of stars
between V$-$I $\approx$ 0.3 and 0.6; subgiant stars evolve quickly 
through this region of the CMD, but at a smoothly increasing rate.
The red edge of the spur at V$-$I $\approx$ 0.3 would imply a sudden 
acceleration in the pace of stellar evolution of low-mass stars which
is not predicted by theoretical models (e.g., Girardi {\it et al.} 1999).
Contrariwise, the color gap is consistent with the location of the 
RR Lyrae instability strip; such gaps are observed in the horizontal branches of
many systems (e.g., the Carina dwarf spheroidal galaxy--- Smecker-Hane
{\it et al.} 1994).

\subsubsection{The Asymptotic Giant Branch}

The strong AGB seen in the WIYN CMD (61 stars redder than V$-$I = 1.8, with
I from 19.2--20.0) is evidence of a large, intermediate-age stellar population
in \ic.  17 of these stars are also present in the WFPC2 CMD.  In the WIYN
image, the density of AGB stars is highest along the bar of the galaxy, 
and decreases with increasing distance from the main body of the galaxy.
The AGB density does not increase towards the areas of current star formation,
northeast of the bar.

Cook, Aaronson \& Norris (1986) developed a relation
between C/M star ratios and heavy element abundances, which was used by
Cook \& Aaronson (1988) to find a metallicity of [Fe/H] $\leq$ $-$1.0 
for the intermediate-age stars in \ic.  Our WIYN I-band AGB luminosity
function resembles the C-star luminosity function 
of Cook {\it et al.} (1986) above I $\approx$ 20, where their sample becomes
incomplete.  Scaling the 14 detected carbon stars from Cook {\it et al.}
to the WIYN field of view predicts $\approx$87 carbon stars; because the 
very red stars are somewhat clustered towards the northwest quadrant of the
WIYN image, where one of the Cook {\it et al.} fields was located, this
overprediction is not a cause for concern.  Conversely, it is not 
unreasonable to expect a large fraction of our 61 AGB stars to be
bona fide carbon stars.  Cook {\it et al.} (1986) found the carbon
star fraction among extended AGB stars to be 75\%--95\% in their two
fields.

Mould \& Aaronson (1982) recognized the fact that 
upper AGB stars are a tracer of 2--6 billion year old stellar populations,
and used the luminosity extent of the AGB as an indicator of stellar age.
Reid \& Mould (1990) examined the AGB population of the Small Magellanic
Cloud and found that, like \ic, the SMC has a high fraction of carbon stars
among the faint ($-$3 $\leq$ M$_{\mathrm bol}$ $\leq$ $-$5.5) AGB stars.
From the AGB luminosity function, Reid \& Mould concluded that the mean
age of the SMC is greater than 2--3 Gyr.  

\begin{figure*}
\centerline{\hbox{\psfig{figure=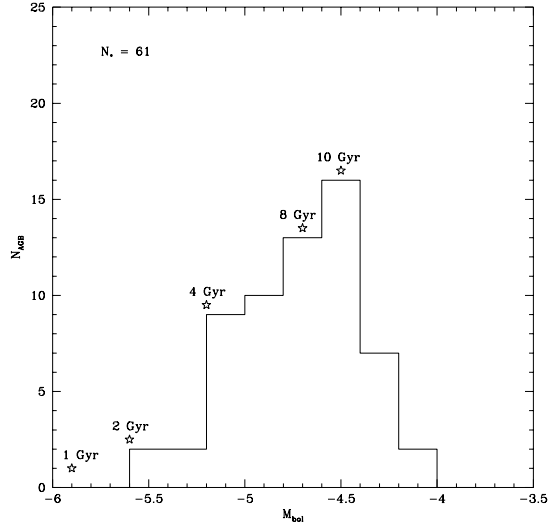,height=3.0in}}}
\caption{Bolometric luminosity function of AGB stars in the WIYN 
field. Results from the carbon star M$_{\mathrm bol}$-age
relation of Mould \& Aaronson (1982) are marked with stars; star
formation has been, on average, a continuous process throughout
the lifetime of \ic, although relative star-formation rates cannot
be accurately inferred from this figure due to model uncertainties
and small number statistics.   \label{cstars}}
\end{figure*}

We perform a similar analysis, adopting the carbon star bolometric
correction from Reid \& Mould (1985),
BC$_{\mathrm I}$ = 1.9 $-$ 0.7(V$-$I)$_0$.
The resulting luminosity function for our WIYN data is shown in 
Figure \ref{cstars}.  The M$_{\mathrm bol}$-age relation from Table
3 of Mould \& Aaronson (1982) is marked with
open stars and labelled according
to age.  Incompleteness is increasing towards
{\it brighter} M$_{\mathrm bol}$:
the most luminous stars in Figure \ref{cstars} are
 located at V $\approx$22.5,
near the limit of the WIYN data.  Therefore we may be biased towards the
detection of somewhat older carbon stars.

It is clear from Figure \ref{cstars} that \ic\ has formed substantial
numbers of AGB stars over a long period of time.  The lack of AGB stars younger
than 2 Gyr is partially due to
the decreasing stellar lifetimes with decreasing age.  However, a 
higher star formation rate during ancient times than during recent ones is
consistent with the presence of the blue horizontal
branch stars.  Information
derived from such an advanced phase of stellar evolution is subject
to compounded uncertainties (see, e.g., Chiosi 1998).  Marigo, Girardi
\& Chiosi (1996), for example, had difficulty reproducing the distribution
of carbon stars in SMC clusters with existing models; even with the latest
models, Marigo \& Girardi (1998) found that the shape of the carbon
star luminosity function is relatively insensitive to the form of the
star-formation history of a galaxy.  Therefore we decline to draw 
quantitative conclusions regarding the variation of star-formation rate
over the past 2--10 Gyr of \ic's history.

\subsubsection{The Red Clump}

The relative strengths of the ancient and middle-aged stellar populations
in dwarf irregular galaxies are still very uncertain (e.g., Tolstoy 1998;
Da Costa 1998; Mateo 1998).
The strong red clump in Figure \ref{vicmd} is an indication of 
the number of 1--10 Gyr old stars.  Because the red clump stars 
share a common core mass (M$_c$ $\approx$ 0.5 M$_{\sun}$), their 
helium-burning lifetimes do not increase rapidly with increasing
age (decreasing total mass) from ages $\approx$ 1.2--10 Gyr.  In 
contrast, the RGB lifetime is a strongly decreasing function of
stellar mass.  The ratio of RC stars to RGB stars in a CMD can
therefore be used as a rough estimate of the mean age of a 
stellar population (see, e.g., 
Han {\it et al.} 1997, Gallagher {\it et al.} 1998, Tolstoy 1998).

In the WFPC2 CMD of \ic, the number ratio of RC to RGB stars
is 2; because photometric completeness at the level of the
red clump (V, I $\approx$ 24.6, 23.8) is $\approx$ (80\%, 90\%),
we take our value N(RC)/N(RGB) $\geq$ 2 to be a
lower limit.  Examination of synthetic CMDs therefore provides an upper
limit to the mean age of the RGB-producing population: 
t$_{\mathrm RGB}$ $\leq$ 7 $\pm$2 Gyr.  The ratio of RC to RGB
lifetimes has a slight metallicity dependence, such that a
more metal-rich population is older than a metal-poor one
at a given N(RC)/N(RGB) (Cole 1999, in preparation).

\begin{figure*}
\centerline{\hbox{\psfig{figure=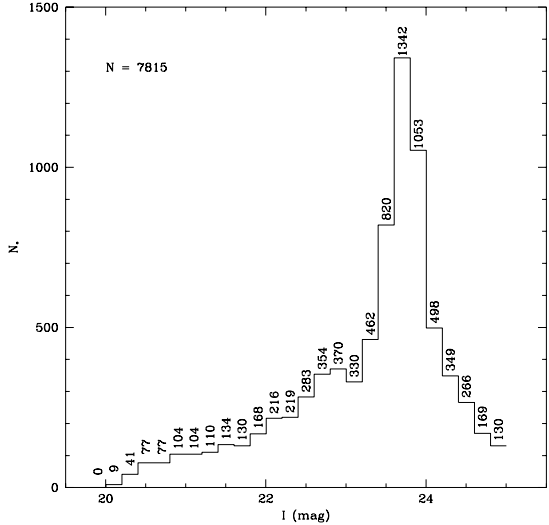,height=3.0in}}}
\caption{RGB I-band luminosity function, highlighting
the red clump.  The mean magnitude of the clump is I = 23.76 $\pm$0.30.
The red clump here contains
a mixture of post-core-helium flash, ``true'' clump stars, 
older than 1.2 Gyr, and
more massive stars aged 0.8--1.2 Gyr.  This combination
and growing incompleteness
at the faint end complicate the interpretation of the clump.
\label{rgblf}}
\end{figure*}

We can make a further estimate of the mean age of the red 
clump stars using its mean I magnitude.  The magnitude of
the red clump depends on its metallicity and, to a lesser
extent, its age (e.g., Girardi {\it et al.} 1998; Seidel,
Demarque, \& Weinberg 1987).  To define the magnitude of 
the red clump, we plot the WFPC2, I-band luminosity function 
in Figure \ref{rgblf}.  The mean magnitude of the red clump
stars is I = 23.76, and the dispersion about the
mean is $\sigma$ = 0.33.  This mean magnitude corresponds to
an absolute magnitude M$_{\mathrm I}$ = $-$0.56.  

If we 
adopt the mean RGB metallicity of [Fe/H] = $-$1.4 for the
red clump, we can then solve for the mean age which produces
a clump of the correct brightness for that metallicity.  
This is most conveniently done relative to the Solar neighborhood,
where M$_{\mathrm I}$(RC) = $-$0.23 $\pm$0.03 (Stanek \& Garnavich 1998).
From Cole (1998), we find that the mean age of the \ic\ red clump
is approximately equal to that of the Solar neighborhood, 
t $\sim$ 6 Gyr (e.g., Jimenez, Flynn, \& Kotoneva 1998).  

Both of these estimates are subject to strong systematic error.
Contamination of the RC by stars in the 800--1200 Myr age range
can bias the measured M$_{\mathrm I}$(RC) either high or low,
depending on the exact star formation and chemical enrichment
histories.  Photometric completeness decreases by a small amount
across the magnitude range of the clump, and may induce a slight
bias toward younger ages.  On a more
fundamental level, there is still some question as to whether
the red clump magnitude varies with age between 2--10 Gyr as
models predict--- Udalski (1998) argues that the clump magnitude
does not vary with age in a sample of Magellanic Cloud star clusters;
however, Girardi (1999) finds that the predicted fading with age
is not ruled out by Udalski's data.  Despite the formiddable 
observational and theoretical uncertainties in the analysis,
we feel that our main conclusion--- that \ic\ is dominated neither
by 10--14 Gyr old stars nor by 1--4 Gyr old stars, but contains stars
of all ages--- is robust.

We can further use the red clump to constrain the metallicities
of the stellar populations of \ic\ at various ages.  The ``pseudo-clump'',
which overlays the red clump but consists of stars just above the maximum
mass of core-helium flash (1.7--2.0 M$_{\sun}$, e.g., Sweigart, Greggio \& 
Renzini 1990) contains stars $\approx$1--1.2 Gyr old.  The color of this
feature has a strong metallicity-dependence (Girardi {\it et al.} 1998);
from the lower limit of V$-$I $\approx$0.75 for the red clump we find
a lower limit of [Fe/H] $\approx$ $-$1.5 for the 1--1.2 Gyr population.

The fading of the red clump with increasing age and metallicity
allows us to put an upper limit on [Fe/H] for the very old stars in \ic,
using the magnitude of the horizontal branch and lower red clump.
For ages older $\approx$6 Gyr, the Z = 0.004 ([Fe/H] = $-$1.3) isochrones
from Girardi {\it et al.} (1999) become too faint to match the observed
red clump in \ic.  This is essentially the same effect that we have seen
for the red horizontal branch stars, which have [Fe/H] $\approx$ $-$1.9.

\section{Discussion}
\label{disc}

The analysis of stellar populations allows us to construct
a rough age-metallicity relation for \ic.  Previous determinations
of \ic's global metallicity have mainly been from nebular spectroscopy
(e.g., Skillman {\it et al.} 1989).  These measures sample only the
youngest component of \ic, and have established the current metal 
abundance of \ic\ at [Fe/H] $\approx$ $-$1.  Freedman (1988a) assessed
the RGB abundance, sampling the stars from 2--14 Gyr old; Cook \& 
Aaronson (1988) judged the upper limit of [Fe/H] to be $\leq$ $-$1 for
the 2--6 Gyr population, using the ratio of C to M stars along 
upper AGB.

\begin{figure*}
\centerline{\hbox{\psfig{figure=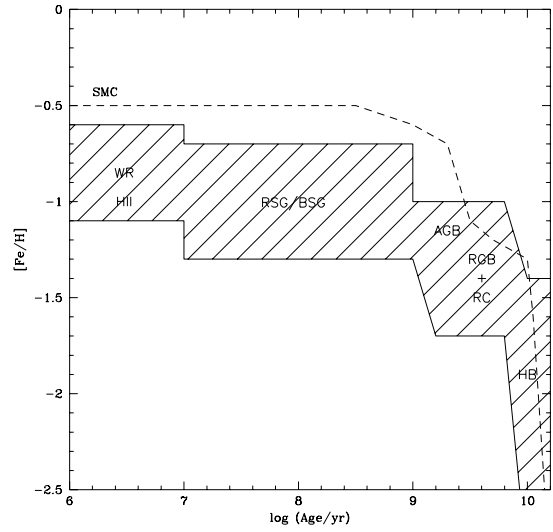,height=3.0in}}}
\caption{Age-metallicity relation for \ic.  Each region of the 
figure is 
labelled with the CMD feature that constrains the metallicity to lie within 
the shaded region.  Abbreviations are as given in the text; WR denotes the 
Wolf-Rayet star; HII denotes the \hii\ regions.  The dotted line shows the
mean age-metallicity relation for the SMC.  \label{amr}}
\end{figure*}

In this paper, we have estimated the metal abundance of the blue and
red supergiants to be [Fe/H] $\approx$ $-$1, providing an abundance 
indicator for stars in the 10--500 Myr range.  We reappraised the
metallicity of the RGB, finding [Fe/H] = $-$1.4 $\pm$0.3.  Using the
morphology of the red clump, we provide an upper limit of [Fe/H]
$\lesssim$ $-$1.3 for the stars older than $\approx$ 6 Gyr.  And
from the magnitude of the horizontal branch stars, we infer a 
metallicity for the oldest stars in \ic\ of [Fe/H] $\sim$ $-$1.9.
We have assembled these data into Figure \ref{amr}, our preliminary
age-metallicity relation for \ic.  Overplotted is an age-metallicity
relation for the Small Magellanic Cloud, from Pagel \& Tautvai\v{s}ien\.{e}
(1998).  Despite the large uncertainties, it appears that 
the chemical evolution of \ic\ has taken a similar form to that of
the SMC, although \ic\ may be slightly more metal-poor.  Figure \ref{amr}
quantifies to an extent the ``population box'' diagram given in Hodge (1989),
although we find a stronger trend of metallicity with age than the 
population box constructed from older data by Grebel (1998).

\begin{figure*}
\centerline{\hbox{\psfig{figure=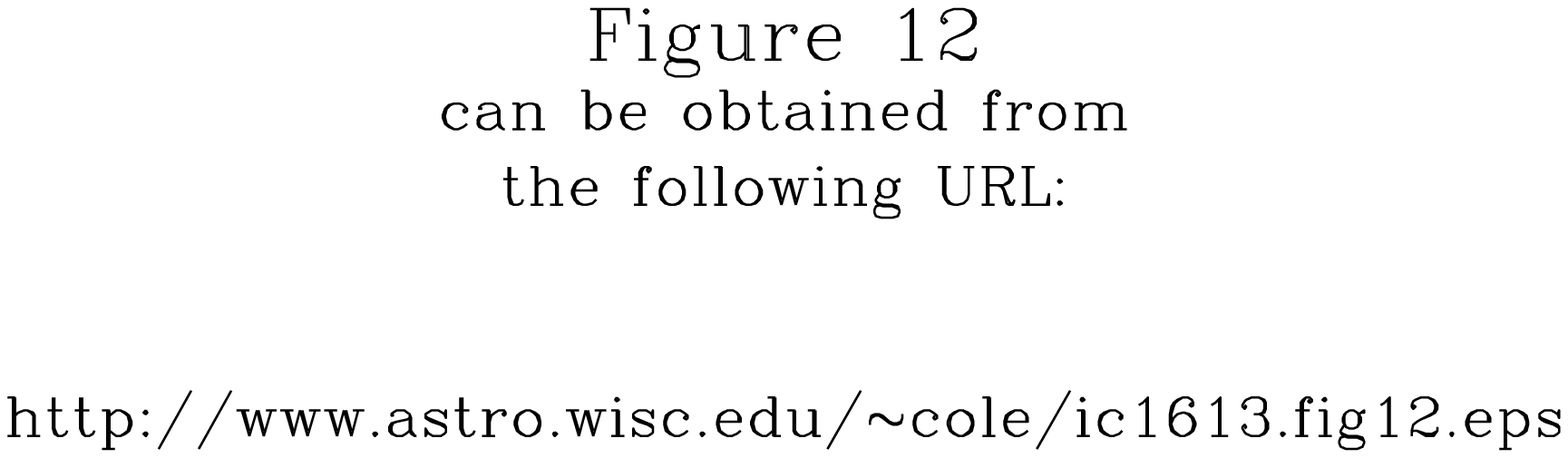,height=5.0in}}}
\caption{Greyscale Hess diagrams for 4 Local Group dwarf galaxies
of similar luminosity: \ic, NGC 147, Pegasus, and Leo A.  The \ic\ logarithmic
Hess diagram is contoured atop each panel to highlight the similarities and
differences between them.   \label{lghess}}
\end{figure*}

Where do \ic's stellar populations place it within the context
of the Local Group dwarf galaxies?  Many Local Group dwarfs have now been studied
with WFPC2 to a comparable depth as \ic, and so we can begin to
use direct intercomparisons to test the accuracy of model-dependent
interpretations of CMDs in terms of galaxy evolution.
In Figure \ref{lghess}, we plot the grayscale Hess diagrams of four
Local Group dwarf galaxies, and overplot the Hess diagram of \ic\ 
for comparison.  The basic parameters of each galaxy,
taken from Mateo (1998), are given in
Table 2.  The quantities $\Delta$(m$-$M) and $\Delta$E$_{\mathrm B-V}$
denote the offsets applied to the data for each galaxy in
order to compare the Hess diagrams to \ic.
Each diagram is constructed from WFPC2 F555W and F814W
data, and plotted with a logarithmic transfer function.  

\begin{deluxetable}{cccccc}
\tablewidth{0pt}
\tablenum{2}
\tablecaption{Four Local Group Dwarfs}
\tablehead{
\colhead{Galaxy} &
\colhead{$\Delta$(m-M)} &
\colhead{$\Delta$E$_{\mathrm B-V}$} &
\colhead{M$_{\mathrm V}$} &
\colhead{[Fe/H]} &
\colhead{M$_{\mathrm HI}$/L$_{\mathrm B}$} 
}
\startdata
IC 1613 & \nodata & \nodata & $-$15.2 & $-$1.4 & 0.81  \nl
NGC 147\tablenotemark{a} & 0.05    & 0.15    & $-$15.5 & $-$1.1 & $<$0.001  \nl
Pegasus\tablenotemark{b} & 0.5     & 0.12    & $-$12.9 & $-$1.3 & 0.44      \nl
Leo A\tablenotemark{c}   & 0.0     & 0.0     & $-$11.4 & $-$1.7 & 1.6 \nl
\enddata
\tablenotetext{a}{Han {\it et al.} 1997.}
\tablenotetext{b}{Gallagher {\it et al.} 1998.}
\tablenotetext{c}{Tolstoy {\it et al.} 1998.}
\end{deluxetable}

\subsection{NGC 147}

NGC 147 is a dwarf elliptical satellite of M31; the WFPC2 data
are from Han {\it et al.} (1997).  Its RGB extends further
to the red than \ic's, due to the higher mean abundance (and/or age) of
NGC 147.  Concordantly, the NGC 147 red clump lies $\approx$0.1
mag below \ic's.  The upper main-sequence is unpopulated
in NGC 147, a sign that star-formation ended at least $\approx$1 Gyr
ago (Han {\it et al.} 1997). 
A blue horizontal branch population is prominent in NGC 147,
sharing its locus in the Hess diagram with that of \ic.  In NGC 147,
star-formation went nearly to completion early in its history, enriching
subsequent generations of stars, which now populate the RGB.  

\subsection{Pegasus}

As mentioned above, Pegasus is a transition galaxy between the 
dwarf irregular and dwarf spheroidal classes; it is also one
of \ic's nearest neighbors within the Local Group.  Though less
massive and less luminous than \ic, Pegasus is of comparable 
metal abundance to \ic, and has a similar RGB and main-sequence
morphology.  However, Pegasus has a larger AGB population than
\ic, as would be expected from its enhanced period of star-formation
a few Gyr ago (Gallagher {\it et al.} 1998).  The recent star-formation
rate in the WFPC2 field, \.{M} $\approx$ 1600 M$_{\sun}$ Myr$^{-1}$
kpc$^{-2}$, is similar to that derived by Gallagher {\it et al.}
(1998) for Pegasus, \.{M} = 1350--2000 M$_{\sun}$ Myr$^{-1}$ kpc$^{-2}$;
however the WFPC2 image of \ic\ excludes the most active regions 
of star formation.  Pegasus and \ic\ may
be examples of galaxies which undergo occasional strong episodes
of star-formation lasting up 0.5--1 Gyr, at intervals of one to a
few Gyr.  The differences in their recent ($<$3 Gyr) star-formation histories
would then be due to a difference in phase in their cycles of quiescent
vs. elevated star-formation.

\subsection{Leo A}

Leo A provides a stark contrast to the other galaxies in this
sample.  The color-magnitude data are from Tolstoy {\it et al.} (1998).
The smallest, most isolated galaxy in the sample, the
CMD of Leo A shows the unusual feature of a dominant blue 
supergiant branch, evidence for strong recent star formation.
Leo A's giant branch lies to the blue of \ic's, indicating its
lower metallicity (and/or age).  The blue, vertically extended red clump 
in Leo A suggests that this galaxy is dominated by an $\approx$
1 Gyr population.  Deep WIYN \& MMT images of Leo A have failed to reveal
an extended Baade's sheet population of red giants that would
be indicative of significant early star formation (A. Saha, private
communication, Tolstoy {\it et al.} 1998).
The low metallicity and high gas fraction
of Leo A are all consistent with a galaxy that has not had a major
star formation episode until quite recently.  Leo A is even more
isolated than \ic\ (its nearest neighbors are the Leo I and Leo II
dwarf spheroidals, more than 450 kpc away), and it must stand
as a counterexample to the Tucana dwarf, a purely old system
(Da Costa 1998), in evolutionary studies of small, isolated galaxies.

\section{Summary}
\label{sum}

We have used WFPC2 to obtain deep images of the dwarf irregular 
galaxy \ic.  We analyze the resulting (I, V$-$I) color-magnitude
diagrams to study the stellar populations at the center of this
object.  The strongest feature of the CMD is the red giant branch
and superposed red clump.  We find a mean metallicity of [Fe/H] = $-$1.4
$\pm$0.3 from the color of the RGB, consistent but slightly lower
than previous results (Freedman 1988).  Upper main-sequence stars are
also present, indicating a continuous period of star-formation 
during the past $\approx$350 Myr at a level of $\sim$ 3.5 $\times$
10$^{-4}$ M$_{\sun}$ yr$^{-1}$ per WFPC2 field, corresponding to
1600 M$_{\sun}$ Myr$^{-1}$ kpc$^{-2}$.  From a comparison of the supergiant
stars to theoretical stellar evolution models, the mean 
chemical abundance during the past 800 Myr
has been approximately one-tenth solar, consistent with previous
measurements of \hii\ region abundances.

In our WIYN image, we find 61 candidate carbon stars which 
populate an extended upper asymptotic giant branch and 
trace intermediate-age (2--10 Gyr) stellar populations.
The AGB luminosity function indicates broadly continuous star formation
from at least 4--10 Gyr ago; a possible decline since then
can be neither confirmed nor ruled out with the present data.

We have detected for the first time a blue horizontal branch 
population in \ic. 
This very old population is not indicative of the \ic\ RGB 
in general:  from an analysis of the red clump stars, we find
a mean age for the RGB of t $\lesssim$ 7 $\pm$2 Gyr.  Systematic
effects are unlikely to push this value upwards by the factor
of $\sim$2 necessary to enforce coevality with the blue horizontal branch.

We present a preliminary age-metallicity relation for \ic.  While
crude, this relation shows the same form as that of the Small
Magellanic Cloud.  We will use the age-metallicity information in 
a detailed modelling of the \ic\ CMD in a subsequent paper.  A 
comparison to WFPC2 CMD's of the galaxies Leo A, NGC 147, and Pegasus
shows that the stellar populations of \ic\ are very similar to 
those of Pegasus.  Pegasus has an excess of bright AGB stars indicating
elevated star formation a few Gyr in the past, while \ic\ shows a higher
level of current star formation; but both galaxies seem to have 
experienced a relatively constant, low level of star formation throughout
their histories.  
Relatively isolated Local Group dwarf galaxies display a considerable
range in their star-formation histories, from objects where star formation
was completed several Gyr in the past (e.g., Tucana--- Lavery {\it et al.}
1996; possibly NGC 147-- Han {\it et al.} 1997),
to young-star dominated systems such as Leo A (Tolstoy {\it et al.} 1998).

\acknowledgments

This research was carried out by the WFPC2 Investigation Definition Team
for JPL and was sponsored by NASA through contract NAS 7--1260.  JSG
and AAC acknowledge travel support from ESO for a visit to Garching
bei M\"{u}nchen.  We would also like to 
thank L\'{e}o Girardi for making the latest isochrones of the Padua group
available in advance of publication. It is a pleasure to thank the WIYN
Telescope support staff for their excellent assistance at Kitt Peak.
AAC thanks Dr. J.C. Howk for 
coffee and suggestions regarding the presentation of Hess diagrams.
This research made use of the NASA Astrophysics Abstract Service, and
the Canadian Astronomy Data Centre, which is operated by the Herzberg
Institute of Astrophysics, National Research Council of Canada.

{}

\newpage

\end{document}